\documentclass[epj]{svjour}

\usepackage{graphicx}
\usepackage{fancyhdr}
\def\makeheadbox{{
 }}
\def\mytitle{My title} 
\def\myauthors{My name}  
\def\mytype{My type of session}
\def\mysession{My session}

\def\mytitle{Instantons in Deformed Super Yang-Mills Theories} 
\def\myauthors{Shin Sasaki}    
\def\mytype{Contributed Talk}    
\def\mysession{Theoretical Models}

\begin{document}
\title{Instantons in Deformed Super Yang-Mills Theories}
\author{Shin Sasaki\inst{1}\thanks{\emph{Email:} shin.sasaki@helsinki.fi}
%
\and Katsushi Ito\inst{2}
\and Hiroaki Nakajima\inst{3}
}                     
%
%
\institute{Department of Physical Sciences, University of Helsinki, P.O.Box 64, Helsinki, Finland
\and
Department of Physics, Tokyo Institute of Technology, Tokyo, 
152-8551, Japan
\and Department of Physics and Institute of Basic Science, Sungkyunkwan University,
Suwon, 440-746, Korea 
}
%
\date{October 10, 2007}
\abstract{
We study the instanton effective action in deformed four-dimensional 
 $\mathcal{N} = 2$ and $\mathcal{N} = 4$ super Yang-Mills (SYM) theories.
These deformed gauge theories are defined on the 
D-brane world-volume in the presence of constant, self-dual 
Ramond-Ramond (R-R) 3-form field strength background $\mathcal{F}$ which is scaled as 
$(2 \pi \alpha')^{\frac{1}{2}} \mathcal{F} = \mathrm{fixed}$ 
in the zero-slope limit $\alpha' \to 0$.
The instanton effective action is 
obtained by solving equations of motion of the deformed $\mathcal{N} = 
2$ SYM action. We show that this effective action correctly reproduces 
the string theory result derived 
from D3/D$(-1)$-branes system in the lowest order of the backgrounds and 
gauge coupling constant. We comment on the generalization of $\mathcal{N} 
= 2$ results to $\mathcal{N} = 4$ case.
} 
\maketitle
\section{Introduction \label{intro}}
Deformation of supersymmetric gauge theories is useful for the 
study of non-perturbative effects. For example, 
the $\Omega$-background and noncommutative deformations of 
$\mathcal{N} = 2$ SYM theory is useful 
 to evaluate integration over the instanton moduli space \cite{Ne}. 
Similar deformations can be provided by introducing supergravity backgrounds.
Especially, constant R-R backgrounds play also an 
interesting role in the instanton calculus.

In fact, it was shown that in D3/D$(-1)$-branes system with 
self-dual graviphoton and vector backgrounds, 
 the deformed effective action of the D$(-1)$-branes 
 is nothing but the instanton effective action of $\mathcal{N} = 2$ 
super Yang-Mills theory in the $\Omega$-background at the lowest 
order in the background \cite{BiFrFuLe}.
Due to the graviphoton and vector backgrounds, the effective action on 
the D-brane world-volume is deformed. We investigated the vacuum structure and 
supersymmetry of the deformed $\mathcal{N} = 2$ and $\mathcal{N} = 4$ SYM actions 
defined on the D3-brane world-volume \cite{ItNaSa}. The instanton effective 
action of these deformed SYM theories should reproduce the result of D3/D$(-1)$-branes
system. In \cite{ItNaSa2}, we derived the instanton equations
in these deformed $\mathcal{N} = 2, \mathcal{N} = 4$ SYM theories and solved it. 
The solutions were obtained through the ADHM construction \cite{AtDrHiMa}.
These solutions are expressed by the ADHM moduli parameters and 
we found that the instanton effective action for the deformed $\mathcal{N} = 
2$ SYM theory agrees with the result from the D3/D$(-1)$-branes system 
obtained in \cite{BiFrFuLe} to first order in the background and 
gauge coupling constant. In the following, we see this 
equivalence in $\mathcal{N} = 2$ case. 
We also comment that it is possible to generalize this result to $\mathcal{N} = 4$ SYM theory.
Detailed results will be explained in \cite{ItNaSa2}.

\section{Deformed gauge theories in constant, self-dual R-R 3-form  background}
To derive the explicit form of the deformed actions on the D3-brane 
world-volume (including fermions), we calculate open string disk amplitudes with 
the insertions of the R-R vertex operators and take the zero-slope limit $\alpha' \to 0$.
The appropriate boundary condition for the vertex operators is 
imposed. We consider the ordinary NSR formalism of type IIB
superstring theory and the R-R background is treated in a perturbative way.
The background $\mathcal{F}^{\alpha \beta AB}$ is scaled as $(2 \pi \alpha')^{\frac{1}{2}} 
\mathcal{F}^{\alpha \beta AB} \equiv C^{\alpha \beta AB}
= \mathrm{fixed}$. Here $\alpha, \beta$ are space-time spinor and $A,B$ 
are internal indices. In addition to this zero-slope scaling, the 
self-duality condition is imposed on the space-time index of the background.
Due to the background, non-trivial interactions are induced 
on the D3-brane world-volume and the supersymmetry which is present 
in the vanishing background case is generically broken.

Let us start from the $\mathcal{N} = 4$ case in the following subsection.
After that, we derive the deformed $\mathcal{N} = 2$ SYM action by 
orbifold projecting the deformed $\mathcal{N} = 4$ action.

\subsection{$\mathcal{N} = 4$ deformed action}
We consider the deformed four-dimensional $\mathcal{N} = 4$ $U(N)$ SYM theory in the presence of 
constant R-R 3-form background $\mathcal{F}^{\alpha \beta AB}$ with the 
self-dual constraint in the four-dimensional space-time part. The deformed action 
is realized on the $N$ Euclidean D3-branes world-volume located in the 
flat $\mathbf{R}^{10}$.

We focus on the leading order corrections of the background, namely, 
the deformed action is obtained by calculating tree-level open string 
amplitudes (disk amplitudes) with the insertion of one R-R closed string vertex operator.

The vertex operators corresponding to open string fields are 
$V_A^{(n)}, V_{\varphi}^{(n)}, V_{\Lambda}^{(n)}, V_{\overline{\Lambda}}
^{(n)}$ which represent a gauge field $A_{\mu}$, adjoint scalar fields 
$\varphi_a \ (a = 1, \cdots 6)$, gauginos $\Lambda^{\alpha A}, 
\overline{\Lambda}_{\dot{\alpha} A}$, $(A = 1, \cdots 4)$ respectively.
Here the notation $(n) \ (n = 0, -1/2, -1)$ stands for the picture number. 
The vertex operators $V_{H_{AA}}^{(0)}$, $V_{H_{A \varphi}}^{(0)}$, 
$V_{H_{\varphi \varphi}}^{(0)}$ for the auxiliary fields \cite{BiFrFuLe}
are also introduced. 
The vertex operator $V_{\mathcal{F}}^{(-1/2,-1/2)}$ 
corresponds to the constant R-R 3-form background.

The following selection rules are useful to find the non-zero amplitudes:
\begin{enumerate}
\item Internal charge cancellation condition among open and closed 
string vertex operators
\item $\alpha'$ power counting
\item Lorentz and internal index structure
\end{enumerate}
The conditions 1 and 3 are necessary for non-vanishing disk amplitudes 
while the condition 2 becomes important when we take the zero-slope limit
$\alpha' \to 0$ and move to the field theory description.
Regarding the condition 2, it is crucial that 
the background R-R field is scaled as $(2 \pi \alpha')^{\frac{1}{2}} 
\mathcal{F} = \mathrm{fixed} = C$ in the zero-slope limit. Considering 
all the conditions 1, 2 and 3, we find that there are only three non-zero 
amplitudes which include one R-R vertex operator:
\begin{eqnarray}
& & \langle \! \langle V^{(0)}_A (p_1) V_{\varphi}^{(-1)} 
(p_2) V^{(-1/2,-1/2)}_{\mathcal{F}} \rangle \! 
 \rangle, \nonumber \\
& & \langle \! \langle V^{(0)}_{H_{AA}} (p_1) V_{\varphi}^{(-1)} 
(p_2) V^{(-1/2,-1/2)}_{\mathcal{F}} \rangle \! 
 \rangle, \nonumber \\
& & \langle \! \langle V^{(-1/2)}_{\Lambda} (p_1) V^{(-1/2)}_{\Lambda} (p_2) V^{(-1/2,-1/2)}_{\mathcal{F}}
\rangle \! \rangle.
\end{eqnarray}
After calculating all the amplitudes, 
taking into account the symmetric factor, taking the zero-slope 
limit and integrating out all the auxiliary fields, 
we find the deformed Lagrangian 
$\mathcal{L}_{\mathcal{N} = 4} = 
\mathcal{L}^{(0)}_{\mathcal{N} = 4}
+\mathcal{L}^{(1)}_{\mathcal{N} = 4}
+\mathcal{L}^{(2)}_{\mathcal{N} = 4} + \cdots$ where 
$\mathcal{L}^{(0)}_{\mathcal{N} = 4}$ is an ordinary 
$\mathcal{N} = 4$ $U(N)$ SYM action and $\mathcal{L}^{(n)}_{\mathcal{N} 
= 4}$ is a correction in $n$-th power of the background.
The first and second order corrections are 
\begin{eqnarray}
\mathcal{L}^{(1)}_{\mathcal{N} = 4}
 &=& \frac{1}{\kappa g^2}
\mathrm{Tr} \left[i F_{\mu \nu} \varphi_a \right] C^{\mu \nu a} 
\nonumber \\
& & - \frac{1}{\kappa g^2}
\mathrm{Tr} \left[ \varepsilon_{ABCD} \Lambda_{\alpha}^{\ A}
	     \Lambda_{\beta}^{\ B}  \right] C^{(\alpha \beta)[CD]},
\label{N41}\\
\mathcal{L}^{(2)}_{\mathcal{N} = 4}
&=& \frac{1}{2} \frac{1}{\kappa g^2} \mathrm{Tr} \left[\varphi_a 
 \varphi_b  \right] C_{\mu \nu}^{\ \ a} C^{\mu \nu b}
\end{eqnarray}
where $\mathrm{Tr} (T^m T^n) = \kappa \delta^{mn}$ for $U(N)$ 
generators $T^m$ and $g$ is a gauge coupling constant.
Here we have defined the deformation parameter by
\begin{eqnarray}
C^{\mu \nu a} &\equiv& - 2 \pi (2 \pi \alpha')^{\frac{1}{2}} (\sigma^{\mu 
 \nu})_{\alpha \beta} \left(\overline{\Sigma}^a \right)_{AB} 
 \mathcal{F}^{(\alpha \beta) [AB]}, \nonumber \\
C^{(\alpha \beta)[AB]} &\equiv& - 2 \pi (2 \pi \alpha')^{\frac{1}{2}}
\mathcal{F}^{(\alpha \beta) [AB]}.
\end{eqnarray}
$\overline{\Sigma}^a$ is a six-dimensional sigma matrix.
Let us check the consistency between our deformed action 
and the known D-brane effective action.
In general, the D-brane effective action is expressed as
\begin{eqnarray}
S = S_{\mathrm{DBI}} + S_{\mathrm{CS}}
\end{eqnarray}
where $S_{\mathrm{DBI}}$ is the Dirac-Born-Infeld action and 
$S_{\mathrm{CS}}$ is a Chern-Simons term.
The R-R background appears in the Chern-Simons term in the 
D-brane effective action
\begin{eqnarray}
S_{\mathrm{CS}} 
&=& \frac{\mu_3}{\kappa} \mathrm{STr} \int_{\mathcal{M}_4}
\! \sum_n P [ e^{i \lambda \mathrm{i}_{\varphi}^{2} }
 \lambda^{\frac{1}{2}} \mathcal{A}^{(n)} ] e^{\lambda F}.
\label{CS}
\end{eqnarray}
The symbol $P$ denotes the pull-back 
of ten-dimensional fields and $\mathrm{i}_{\varphi}$ is 
the interior product by $\varphi_a$. $\mu_3$ is a R-R 
charge of the D3-brane, $\lambda = 2 \pi \alpha'$ and STr is a symmetric trace of the $U (N)$ gauge group.
Finally, $\mathcal{A}^{(n)}$ is an $n$-form R-R potential.

It is not difficult to check that the linear term in $C$ in the 
bosonic part of the deformed action (\ref{N41}) 
precisely agrees with the eq. (\ref{CS}).
Since our deformation is due to the R-R background,
it would be expected that there is a fuzzy sphere 
configuration of the vacuum. Indeed, we showed that 
there is a fuzzy $S^2$ solution of the vacuum in the 
deformed $\mathcal{N} = 4$ SYM theory. 
For more detail, see \cite{ItNaSa}.

\subsection{$\mathcal{N} = 2$ deformed action}
The deformed $\mathcal{N} = 2$ SYM action can be derived by 
orbifold projecting the deformed $\mathcal{N} = 4$ action 
obtained in the previous subsection.
The $\mathcal{N} = 2$ $U(N)$ SYM theory is realized on the $N$ 
(fractional) D3-branes located in the orbifold fixed point.
Directions perpendicular to the D3-branes are orbifolded as 
$\mathbf{C} \times \mathbf{C}^2/\mathbf{Z}_2$. 
The vertex operators are 
appropriately projected and the surviving open string degrees of freedom 
are $A_{\mu}$, $\varphi$, $\bar{\varphi}$ corresponding to gauge field, adjoint 
scalars, and $ \Lambda^{\alpha I}$, $\overline{\Lambda}_{\dot{\alpha}I}
 \ (I = 1,2)$ corresponding to gauginos. The R-R background is now 
decomposed into the form 
\begin{eqnarray}
C^{\alpha \beta AB} = - \frac{i}{2 \sqrt{2}} \left(
\begin{array}{cc}
C^{\alpha \beta} & 0 \\
0 & - \bar{C}^{\alpha \beta} 
\end{array}
\right)
\end{eqnarray}
by the orbifold projection. Here $C^{\alpha \beta}$ is a
graviphoton and $\bar{C}^{\alpha \beta}$ is a vector background.
Both of these satisfy the self-duality condition.
By this projection, the $\mathcal{N} = 2$ deformed Lagrangian 
$\mathcal{L}_{\mathcal{N} = 2} = \mathcal{L}^{(0)}_{\mathcal{N} = 2} 
+ \mathcal{L}^{(1)}_{\mathcal{N} = 2} + \mathcal{L}^{(2)}_{\mathcal{N} = 2} 
+ \cdots $ is obtained from the deformed $\mathcal{N} = 4$ SYM Lagrangian. 
Here $\mathcal{L}^{(0)}_{\mathcal{N} = 2}$ 
is an ordinary $\mathcal{N} = 2$ SYM Lagrangian and 
\begin{eqnarray}
& & \mathcal{L}^{(1)}_{\mathcal{N} = 2}
+\mathcal{L}^{(2)}_{\mathcal{N} = 2} =
\frac{1}{\kappa g^2_{\mathrm{YM}} } \mathrm{Tr} \biggl[
i(C^{\mu\nu}\bar{\varphi}+\bar{C}^{\mu\nu}\varphi)
F_{\mu\nu} \nonumber \\
& & \qquad -\frac{1}{\sqrt{2}}\bar{C}^{\mu\nu}
\Lambda^{I}\sigma_{\mu\nu}\Lambda_{I}
+\frac{1}{2}
(C^{\mu\nu}\bar{\varphi}+\bar{C}^{\mu\nu}\varphi)^{2}
\biggr]. \label{N2_correction}
\end{eqnarray}
The supersymmetry of this deformed action is generically broken 
due to the background. However, depending on the rank of the 
background, there are unbroken deformed supersymmetries 
in this model. The number of remaining supersymmetries is summarized in 
table \ref{n2susy}.
\begin{table}[htbp]
\caption[tablesusy2]{The number of unbroken supersymmetries in 
the deformed $\mathcal{N}=2$ SYM action \cite{ItNaSa}. $R[M]$ is the rank of the matrix $M$.}
\label{n2susy}
\begin{tabular}{|c|c|l|l|l|}
\hline
\multicolumn{2}{|c|}{ } & 
\multicolumn{3}{|c|}{$R[C^{(\alpha\beta)}$]}\\
\cline{3-3}\cline{4-4}\cline{5-5}\multicolumn{2}{|c|}{ } & 
\multicolumn{1}{|c|}{0} & \multicolumn{1}{|c|}{1} & \multicolumn{1}{|c|}{2} \\
\hline
  & 0 & $\mathcal{N}=(1,1)$ & $\mathcal{N}=(1,0)$ & $\mathcal{N}=(1,0)$ \\
\cline{2-5}
$R[\bar{C}^{(\alpha\beta)}]$ & 1 & $\mathcal{N}=(\frac{1}{2},1)$ & 
$\mathcal{N}=(\frac{1}{2},0)$ & $\mathcal{N}=(\frac{1}{2},0)$ \\
\cline{2-5}
 & 2 & $\mathcal{N}=(0,1)$ & $\mathcal{N}=(0,0)$ & $\mathcal{N}=(0,0)$ \\
\hline
\end{tabular}
\end{table}
For more detail, see the reference \cite{ItNaSa}.

\section{Instanton calculus in deformed super Yang-Mills theories}
Let us focus on the instanton effective action in the deformed 
$\mathcal{N} = 2$ SYM theory. To derive the instanton effective action,
 we need to find the instanton solution which is expressed by the 
ADHM moduli parameters \cite{DoHoKhMa}.
After writing the action to the perfect square form, the gauge field 
strength part and the last term in (\ref{N2_correction}) become
\begin{eqnarray}
\mathcal{L}_{\mathrm{gauge}} &=& 
\frac{1}{\kappa} \mathrm{Tr} \left[
- \frac{1}{2} (F_{\mu \nu}^{(-)})^2 \right. 
\nonumber \\
& & \left. - \frac{1}{2} \left( 
F_{\mu \nu}^{(+)} - i g (C^{\mu \nu} \bar{\varphi} + \bar{C}^{\mu \nu} \varphi)
\right)^2 \right],
\end{eqnarray}
where $F^{(\pm)}_{\mu \nu} = \frac{1}{2} (F_{\mu \nu} \pm \tilde{F}_{\mu 
\nu})$, $\tilde{F}_{\mu \nu} = \frac{1}{2} \varepsilon_{\mu \nu \rho 
\sigma} F^{\mu \nu}$.
The (anti)instanton equation is derived from 
$\mathcal{L}_{\mathrm{gauge}}$ as 
\begin{eqnarray}
\!\!\! & & \!\!\! F^{(-)} = 0 \ \textrm{(instanton)}, 
\nonumber \\
\!\!\! & & \!\!\! F^{(+)}_{\mu \nu} - i g (C^{\mu \nu} \bar{\varphi} + \bar{C}^{\mu \nu} 
\varphi) = 0 \ \textrm{(anti-instanton)}.
\end{eqnarray}
Since we are interested in the low-energy effective theory 
of the $\mathcal{N} = 2$ pure SYM theory, 
we introduce the VEV of the adjoint fields $\varphi, \bar{\varphi}$.
However, it is known that when the adjoint scalar fields have a VEV, 
the super instanton solution is expanded in the gauge coupling constant $g$ and 
the solution is determined in a perturbative way. Let us focus on the 
self-dual condition. In this case, the gauge coupling expansion 
of the solution is
\begin{eqnarray}
A_{\mu} &=& g^{-1} A^{(0)}_{\mu} + 
g A_{\mu}^{(1)} + \cdots, \nonumber \\
\Lambda^I &=& g^{- \frac{1}{2}} \Lambda^{(0)I} + 
g^{\frac{3}{2}} \Lambda^{(1)I} + \cdots, \nonumber \\
\bar{\Lambda}_I &=& g^{\frac{1}{2}} \bar{\Lambda}^{(0)}_I
+ g^{\frac{5}{2}} \bar{\Lambda}^{(1)}_I + \cdots, \nonumber \\
\varphi &=& g^0 \varphi^{(0)} + g^2 \varphi^{(1)} + \cdots, 
\nonumber \\
\bar{\varphi} &=& g^0 \bar{\varphi}^{(0)} + g^2 
\bar{\varphi}^{(1)} + \cdots.
\end{eqnarray}
The equation of motion up to leading order in $g$ is now
\begin{eqnarray}
& & F^{(0)(-)}_{\mu \nu} = 0, 
\nonumber \\
& & \nabla^2 \bar{\varphi}^{(0)} + i F^{(0)}_{\mu \nu} \bar{C}^{\mu \nu}= 0, 
\nonumber \\
& & \nabla^2 \varphi^{(0)} + i \sqrt{2}  \Lambda^{(0)I} 
\Lambda^{(0)}_I + i F^{(0)}_{\mu \nu} C^{\mu \nu} = 0, 
\nonumber \\
& & (\sigma^{\mu})_{\alpha \dot{\beta}} \nabla_{\mu}
\overline{\Lambda}^{(0)}_I {}^{\dot{\beta}} + \sqrt{2} 
[\bar{\varphi}^{(0)}, \Lambda^{(0)}_{I \alpha}] 
+ \sqrt{2} \Lambda^{(0)\beta} {}_I \bar{C}_{(\beta \alpha)}= 0, 
\nonumber \\
& & (\bar{\sigma}^{\mu})^{\dot{\alpha} \beta} \nabla_{\mu}
\Lambda^{(0)I}_{\beta} = 0, 
\nonumber \\
& & \nabla_{\mu} (F^{(0) \mu \nu} + \tilde{F}^{(0) \mu \nu}) = 0.
\end{eqnarray}
Here $\nabla_{\mu}$ is a gauge covariant derivative defined by zeroth order 
gauge field $A_{\mu}^{(0)}$. $F_{\mu \nu}^{(0)}$ is a field strength 
of $A_{\mu}^{(0)}$.
It is easy to see that the self-duality condition $F^{(0)(-)}_{\mu \nu} = 0$ is consistent with the 
equation of motion. The solution to this equation can be expressed 
by the ADHM moduli parameters $(a'_{\mu}, \mathcal{M}', \mu, \bar{\mu}, 
\chi, \bar{\chi}, w, \bar{w})$ \cite{ItNaSa2}. 
First of all, the gauge field equation is not deformed by the background.
Thus the gauge field part is an ordinary $k$ instanton background. 
For the same reason, the solution for $\Lambda$ is not deformed.
On the other hand, $\overline{\Lambda}$ and adjoint scalar equations in this 
background is deformed by the non-zero background\footnote{We 
 need not to solve the equation of motion for $\overline{\Lambda}$ 
because $\overline{\Lambda}$ contributes to the classical potential 
through the sub-leading order in $g$.}.
The result is 
\begin{eqnarray}
& & A^{(0)}_{\mu} = -i \overline{U} \partial_{\mu} U, \nonumber \\
& & \Lambda^{(0)I}_{\alpha} = \Lambda_{\alpha} (\mathcal{M}^I)
= \overline{U} (\mathcal{M}^I f \bar{b}_{\alpha} - b_{\alpha} f 
\overline{\mathcal{M}}^I) U, \nonumber \\
& & \varphi^{(0)} = i \frac{\sqrt{2}}{4} \epsilon_{IJ}
\overline{U} \mathcal{M}^I f \overline{\mathcal{M}}^J U + \overline{U}
\left(
\begin{array}{cc}
\phi & 0 \\
0 & \chi \mathbf{1}_{2} + \mathbf{1}_{k} C
\end{array}
\right)
U, \nonumber \\
& & \bar{\varphi}^{(0)} = 
\overline{U} \left(
\begin{array}{cc}
\bar{\phi} & 0 \\
0 & \bar{\chi} \mathbf{1}_{2} + \mathbf{1}_{k} \bar{C}
\end{array}
\right)
U,
\label{N2_solution}
\end{eqnarray}
where $C_{\alpha} {}^{\beta} = (\sigma^{\mu \nu})_{\alpha} {}^{\beta} 
C^{\mu \nu}$, $\mathcal{M} = (\mu, \mathcal{M}')^{T}$ 
and $\phi = \langle \varphi^{(0)} \rangle, \ \bar{\phi} = \langle 
\bar{\varphi}^{(0)} \rangle$ are VEVs of the adjoint scalars. 
$U_{\lambda u}$ is an $(N + 2 k) \times N$ matrix which satisfies the conditions
\begin{eqnarray}
& & \overline{\Delta}_i {}^{\dot{\alpha} \lambda} U_{\lambda u} = 0, 
\quad \overline{U}_u {}^{\lambda} U_{\lambda u} = \delta_{uv}, \nonumber \\
& &  \Delta_{\lambda i \dot{\alpha}} f_{ij} \overline{\Delta}_j 
{}^{\dot{\alpha} \rho} = \delta_{\lambda} {}^{\rho} - U_{\lambda u} 
\overline{U}^{u \rho}, \nonumber \\
& & \Delta_{\lambda j \dot{\alpha}} (x) \equiv \Delta_{(u + i \alpha)j\dot{\alpha}}
= \left(
\begin{array}{c}
w_{u j \dot{\alpha}} \\
\delta_{ij} x_{\alpha \dot{\alpha}} + 
(a'_{\alpha \dot{\alpha}})_{ij}
\end{array}
\right).
\end{eqnarray}
Here we have introduced the indices $\lambda = 1, 2, \cdots, N + 2k$, $i, j = 
1, 2, \cdots, k$, $u, v = 1, 2, \cdots, N$. $f$ is an $x$-dependent $k 
\times k$ matrix. $\chi, \bar{\chi}$ 
should satisfy the constraints
\begin{eqnarray}
& & \mathbf{L} \chi = 
- i \frac{\sqrt{2}}{4} \epsilon_{IJ}
\overline{\mathcal{M}}^I \mathcal{M}^J 
+\bar{w}^{\dot{\alpha}} \phi w_{\dot{\alpha}}
+ C^{\mu \nu} [a'_{\mu}, 
a'_{\nu}], \nonumber \\
& & \mathbf{L} \bar{\chi} = 
\bar{w}^{\dot{\alpha}} \bar{\phi} w_{\dot{\alpha}}
+ \bar{C}^{\mu \nu} [a'_{\mu}, a'_{\nu}] \label{chi_const},
\end{eqnarray}
where the operator $\mathbf{L}$ is defined by
\begin{eqnarray}
\mathbf{L} * = \frac{1}{2} \{ \bar{w}^{\dot{\alpha}} w_{\dot{\alpha}}, * \}
+ [a'_{\mu}, [a'^{\mu}, *]].
\end{eqnarray}
The classical action is now expanded as 
\begin{eqnarray}
S = \frac{8 \pi^2 k}{g^2} + i k \theta + g^0 S^{(0)}_{\mathrm{eff}} +
\mathcal{O} (g^2), \label{classical_action_expansion}
\end{eqnarray}
where we have introduced the $\theta$ angle. The 
leading order action is 
\begin{eqnarray}
S^{(0)}_{\mathrm{eff}} &=& 
\frac{1}{\kappa} \int \! d^4 x \ \mathrm{Tr} \left[ 
- \nabla_{\mu} \varphi^{(0)} \nabla^{\mu} \bar{\varphi}^{(0)} \right. 
\nonumber \\
& & - \frac{i}{\sqrt{2}}
\Lambda^{(0)I} [\bar{\varphi}, \Lambda^{(0)}_I]
 + i \bar{\varphi}^{(0)} F^{(0)}_{\mu \nu} C^{\mu \nu}
\nonumber \\
& & \left. + i \varphi^{(0)} 
F^{(0)}_{\mu \nu} \bar{C}^{\mu \nu} 
+ i \frac{\sqrt{2}}{2}   \Lambda_{\alpha}^{(0)I} \Lambda_{\beta I}^{(0)} \bar{C}^{(\alpha \beta)}
\right]. \nonumber \\
\label{classical_n2}
\end{eqnarray}
After plugging the solution (\ref{N2_solution}) into the eq. (\ref{classical_n2}), 
the instanton effective action is evaluated as
\begin{eqnarray}
S^{(0)}_{\mathrm{eff}} &=&
4\pi^{2}\mathrm{tr}_{k}\biggl[-i\frac{\sqrt{2}}{4}\epsilon_{IJ}
\bar{\mu}^{I}\bar{\phi}\mu^{J}+\frac{1}{2}\bar{w}^{\dot{\alpha}}
(\bar{\phi}\phi+\phi\bar{\phi})w_{\dot{\alpha}}
\nonumber \\
& & -\bar{\chi}\mathbf{L}^{-1}\chi
-i\frac{\sqrt{2}}{8}\bar{C}^{(\alpha\beta)}\epsilon_{IJ}
\mathcal{M}^{\prime I}_{\alpha}\mathcal{M}^{\prime J}_{\beta}
\biggr] \nonumber \\
& & +(\bar{C}C\mbox{-term}).
\end{eqnarray}
It is possible to rewrite the result in \cite{BiFrFuLe} in the above 
form just by imposing an ADHM constraint, dropping the 
Lagrange multiplier term in \cite{BiFrFuLe} and solving
$\chi, \bar{\chi}$ in terms of other moduli. We can see that our 
result precisely reproduces the known result which was obtained in 
D3/D$(-1)$-brane system in the presence of self-dual R-R background 
up to the $\mathcal{O} (C \bar{C})$ terms. 
Thus our deformed action reproduces the result 
which is consistent with string theory calculations.

Let us next consider the anti-self-dual configuration 
$F^{(+)} = 0$. The solution is expanded as 
\begin{eqnarray}
A_{\mu} &=& g^{-1} A^{(0)}_{\mu} + 
g A_{\mu}^{(1)} + \cdots, \nonumber \\
\Lambda^I &=& g^{ \frac{1}{2}} \Lambda^{(0)I} + 
g^{\frac{5}{2}} \Lambda^{(1)I} + \cdots, \nonumber \\
\bar{\Lambda}_I &=& g^{- \frac{1}{2}} \bar{\Lambda}^{(0)}_I
+ g^{\frac{3}{2}} \bar{\Lambda}^{(1)}_I + \cdots, \nonumber \\
\varphi &=& g^0 \varphi^{(0)} + g^2 \varphi^{(1)} + \cdots, \nonumber \\
\bar{\varphi} &=& g^0 \bar{\varphi}^{(0)} + g^2 
\bar{\varphi}^{(1)} + \cdots.
\end{eqnarray}
The leading order equation is then
\begin{eqnarray}
& & F^{(0)(+)}_{\mu \nu} = 0, \nonumber \\
& & \nabla^2 \bar{\varphi}^{(0)} - i \sqrt{2} 
\overline{\Lambda}^{(0)}_I \overline{\Lambda}^{(0)I} = 0, \nonumber \\
& & \nabla^2 \varphi^{(0)} = 0, \nonumber \\
& & (\sigma^{\mu})_{\alpha \dot{\beta}} \nabla_{\mu} 
\overline{\Lambda}^{(0) \dot{\beta}}_I = 0, \nonumber \\
& & (\bar{\sigma}^{\mu})^{\dot{\alpha} \beta} \nabla_{\mu} 
\Lambda{(0)I}_{\beta} - \sqrt{2} [\varphi^{(0)}, 
\overline{\Lambda}^{(0) I \dot{\alpha}}] = 0, \nonumber \\
& & \nabla_{\mu} (F^{(0) \mu \nu} + \tilde{F}^{(0) \mu \nu}) = 0.
\end{eqnarray}
Thus the equation of motion does not receive 
corrections from $C, \bar{C}$. Moreover, $S^{(0)}_{\mathrm{eff}}$ in 
(\ref{classical_action_expansion}) for the anti-self-dual gauge 
coupling expansion is 
\begin{eqnarray}
S^{(0)}_{\mathrm{eff}} &=& 
\frac{1}{\kappa} \int \! d^4 x \ \left[
- \nabla_{\mu} \varphi^{(0)} \nabla^{\mu} \bar{\varphi}^{(0)}
+ \frac{i}{\sqrt{2}} \overline{\Lambda}_I^{(0)} [\varphi^{(0)}, \overline{\Lambda}^{(0)I} ]
\right.
\nonumber \\
& & \left. + i F_{\mu \nu}^{(0)} \bar{\varphi}^{(0)} C^{\mu \nu} 
+ i F_{\mu \nu}^{(0)} \varphi^{(0)} \bar{C}^{\mu \nu} 
\right].
\end{eqnarray}
Due to the self-duality condition on the background $C, \bar{C}$,
there are no background corrections in the instanton effective action.
Therefore we conclude that the anti-self-dual sector is not deformed by 
the self-dual background.

The super instanton solution in the $\mathcal{N} = 4$ deformed SYM 
theory is also obtained in \cite{ItNaSa2}.
The deformed instanton equation is derived in a straightforward way and
the solution to the equation is expressed by the ADHM moduli parameters 
as in the case of $\mathcal{N} = 2$ deformed SYM theory. By using this solution, we can 
calculate the instanton effective action in principle.

However, there is an efficient way to derive the deformed $\mathcal{N} = 4$ 
instanton effective action. Since it is known that the $\mathcal{N} = 2$
deformed SYM action is obtained from the $\mathcal{N} = 4$ theory by 
 the orbifold projection, the $\mathcal{N} = 4$ instanton effective action 
is derived from the the consistency of $\mathcal{N} = 2$ and $\mathcal{N} 
= 4$ instanton effective actions by the same orbifold projection. 
The result can be found in \cite{ItNaSa2}.

These deformed gauge theories and the deformed instanton effective 
actions have very similar structure to the $\Omega$-background 
deformation \cite{BiFrFuLe}. 
It would be expected that these deformation effects play a role similar to 
the $\Omega$-background in the instanton calculus. 
For more detail, see the reference \cite{ItNaSa2} and future works.
\\
\\
\noindent {\large \bf Acknowledgements}
\\
\\
S.~S. would like to thank C.~Montonen for careful reading of the 
manuscript. 
The talk of S.~S. in this conference is supported by 
Inoue foundation for science. The work of S.~S.
is supported by the bilateral program of Japan Society
for the Promotion of Science (JSPS) and Academy of Finland, ``Scientist
Exchanges.''

\end{document}